\documentclass[10pt,conference]{IEEEtran}

\usepackage{graphicx}

\usepackage[caption=false]{subfig}
\usepackage{xcolor}
\usepackage{balance}
\usepackage[noadjust]{cite}
\usepackage{url}
\usepackage{booktabs}
\usepackage{amsfonts}
\usepackage{amsmath}
\usepackage{hyperref}
\usepackage{cleveref}
\usepackage{listings}

\AtBeginDocument{%
  }

\usepackage{soul}
\usepackage{makecell}
\usepackage{multirow}
\newcommand{\lessonlearned}[1]{\begin{center}
		\framebox{
			\begin{minipage}{0.93\columnwidth}
				{} \textit{#1}
			\end{minipage}
		}
	\end{center}
}

\makeatletter
\newcommand{\linebreakand}{%
  \end{@IEEEauthorhalign}
  \hfill\mbox{}\par
  \mbox{}\hfill\begin{@IEEEauthorhalign}
}
\makeatother

\begin{document}

\title{Towards Realistic Evaluation of Commit Message Generation by Matching Online and Offline Settings}




\author{
\IEEEauthorblockN{Petr Tsvetkov*}
\IEEEauthorblockA{\textit{JetBrains Research}\\
Bremen, Germany\\
petr.tsv@gmail.com}

\and

\IEEEauthorblockN{Aleksandra Eliseeva*}
\IEEEauthorblockA{\textit{JetBrains Research}\\
Belgrade, Serbia\\
alexandra.eliseeva@jetbrains.com}

\and

\IEEEauthorblockN{Danny Dig}
\IEEEauthorblockA{\textit{JetBrains Research}\\
\textit{University of Colorado}\\
Colorado, USA\\
danny.dig@jetbrains.com}

\linebreakand

\IEEEauthorblockN{Alexander Bezzubov}
\IEEEauthorblockA{\textit{JetBrains}\\
Amsterdam, The Netherlands\\
alexander.bezzubov@jetbrains.com}
\and

\IEEEauthorblockN{Yaroslav Golubev}
\IEEEauthorblockA{\textit{JetBrains Research}\\
Belgrade, Serbia\\
yaroslav.golubev@jetbrains.com}

\and

\IEEEauthorblockN{Timofey Bryksin}
\IEEEauthorblockA{\textit{JetBrains Research}\\
Limassol, Republic of Cyprus\\
timofey.bryksin@jetbrains.com}

\linebreakand

\IEEEauthorblockN{Yaroslav Zharov}
\IEEEauthorblockA{\textit{JetBrains Research}\\
Munich, Germany\\
yaroslav.zharov@jetbrains.com}

}

\maketitle

\newcommand{\todo}[1]{\hl{\textbf{TODO: ``#1''}}}

\begin{abstract}

Commit message generation (CMG) is a crucial task in software engineering that is challenging to evaluate correctly. When a CMG system is integrated into the IDEs and other products at JetBrains, we perform \textit{online} evaluation based on user acceptance of the generated messages. However, performing online experiments with every change to a CMG system is troublesome, as each iteration affects users and requires time to collect enough statistics. On the other hand, \textit{offline} evaluation, a prevalent approach in the research literature, facilitates fast experiments but employs automatic metrics that are not guaranteed to represent the preferences of real users. 
In this work, we describe a novel way we employed to deal with this problem at JetBrains, by leveraging an online metric---the number of edits users introduce before committing the generated messages to the VCS---to select metrics for offline experiments.

To support this new type of evaluation, we develop a novel markup collection tool mimicking the real workflow with a CMG system, collect a dataset with 57 pairs consisting of commit messages generated by GPT-4 and their counterparts edited by human experts, and design and verify a way to synthetically extend such a dataset. Then, we use the final dataset of 656 pairs to study how the widely used similarity metrics correlate with the online metric reflecting the real users' experience.

Our results indicate that edit distance exhibits the highest correlation with the online metric, whereas commonly used similarity metrics such as BLEU and METEOR demonstrate low correlation. This contradicts the previous studies on similarity metrics for CMG, suggesting that user interactions with a CMG system in real-world settings differ significantly from the responses by human labelers within controlled environments. 
While our findings are tied to the CMG model we used, and the results may vary for models with cardinally different outputs, our proposed framework is relatively lightweight in terms of required human effort, and we release all the code and the dataset to support future research in the field: \url{https://jb.gg/cmg-evaluation}.

\end{abstract}


\section{Introduction}

\renewcommand*{\thefootnote}{\fnsymbol{footnote}}
\footnotetext[1]{\vspace{-0.2cm}These authors contributed equally to the work.}
\renewcommand*{\thefootnote}{\arabic{footnote}}

Given the extensive use of Version Control Systems (VCS) in software projects, developers daily face the necessity to write \textit{commit messages}---short natural language comments accompanying code changes.
Since developers may neglect manually writing them due to time and effort constraints~\cite{good-commit-message, empty_msgs}, \textit{commit message generation (CMG)} is both a widely studied research topic~\cite{tao2021evaluation, zhang2024automatic} and a prominent feature in IDEs and other products offering software engineering assistance, \emph{e.g.,} Microsoft Copilot~\cite{copilot} or JetBrains AI Assistant~\cite{aiassistant}.

A critical aspect of developing a CMG system is evaluating its performance. Ideally, an evaluation metric should align with how software developers perceive the system's quality. While \textit{human offline evaluation}---collecting quality labels for tool-generated commit messages from paid assessors---precisely achieves this, it is very resource-intensive. Therefore, both academia and industry search for a more practical way to assess the quality of the CMG system. 

The prevalent approach in research is \textit{reference-based offline evaluation}---calculating an automatic text similarity metric or a set of metrics between the original commit message from the VCS (we refer to it as $O$) and the tool-generated commit message $G$~\cite{tao2021evaluation}. A key challenge for this evaluation setup is the selection of the text similarity metric to ensure that the results are aligned with human perception, as suggested by numerous research works on similarity metrics for CMG~\cite{tao2021evaluation, hu2022correlating, zhang2024usinglargelanguagemodels}.

Industry-grade CMG systems offer a practical method to assess human preferences through \textit{online evaluation}---the real-time assessment of a system’s performance using live user interactions and feedback. The metrics might include whether users accepted the generated commit message, to what extent the users modified the generated commit message before committing, and more. However, online evaluation complicates fast experimentation since it requires deploying the system and collecting the user logs for a period of time. When designing AI-assisted features for IDEs and other tools for software developers at JetBrains, we needed a faster alternative suitable for iterative experiments with CMG models and embarked on a journey to find suitable metrics for the reference-based offline evaluation that aligns with our needs.

Although previous studies have investigated how well the automatic text similarity metrics for CMG align with human preferences, these works have been limited to controlled experimental settings using fixed questions designed by researchers. As a result, it remains unclear whether these metrics accurately reflect the performance of CMG models in real-world production environments with numerous users that may have diverse perceptions of commit message quality.

In this work, we bridge this gap and present a pipeline to select the best metric for the reference-based offline evaluation given the target online metric. The heart of our framework is the observation that in real-world CMG systems deployed in production, users typically edit the generated commit messages $G$ before committing them to the VCS, thus introducing another entity---edited messages $E$. 

Building off this idea, we propose to employ multiple pairs of generated messages $G_i$ and their edited versions $E_i$ for each commit $C$ in the offline metric selection process. Each \textit{related pair}, where $E_i$ was obtained directly from $G_i$, provides us with an intuitive and practical metric---the effort that the editing process took---thus providing the online evaluation of the $G_i$. At the same time, edited messages $E_{j\neq i}$ obtained from other generated messages for the same commit, as well as the original commit message $O$ scraped from the VCS, provide reference messages for $G_i$, suitable for offline evaluation. Hence, we can evaluate each generated message $G_i$ both online (against $E_i$) and offline (against $E_{j\neq i}$ and $O$). This allows us to gauge how good different metrics for the offline evaluation will estimate the online performance of the model.

For this work, we select the classic edit distance (ED)~\cite{levenshtein} metric as the primary online metric in our experiments because it reflects the amount of work users put in before accepting the proposed message. The ideal CMG system should provide messages that users can accept without editing. For offline metrics, we consider 9 text similarity metrics commonly used in CMG research. Our findings reveal that for our model, only edit distance and edit similarity (a normalized version of edit distance) exhibit a high or moderate correlation with online ED values, while the remaining metrics show low correlations.

These results do not align with the previous findings based on human annotations via a fixed question rubric. For instance, Tao et al.~\cite{tao2021evaluation} show that the best variant of BLEU metric achieves $0.62$ Spearman correlation with a 5-point Likert scale human scores, while in our setup BLEU achieves only a low Spearman correlation of $-0.17$. Similarly, Hu et al.~\cite{hu2022correlating} asked the human annotators to label commit messages using several different criteria (\emph{e.g.}, Naturalness, Conciseness, Usefulness). Among the automatic text similarity metrics, METEOR~\cite{banarjee2005} achieves the highest Kendall correlation of $0.5$ with these scores, while our experiments show a very low Spearman correlation of $0.04$. 
These discrepancies suggest that correlations with the scores obtained from the controlled surveys of human annotators may not accurately reflect the relevance of offline metrics to online performance in particular production settings. With this work, we want to emphasize the importance of applied research for industry products and share the practical experience of such research.

As part of this effort, we developed an app for collecting the expert markup for the generated commit messages and used it to extend a CMG dataset from Long Code Arena~\cite{lca} with multiple $(G,E)$ pairs per commit. For 15 commit messages generated by GPT-4~\cite{gpt4}, we collected 57 edits from human experts and extended the dataset by synthetically generating 599 more $(G,E)$ pairs. We assess the distribution of our dataset in relation to logs from real users of a CMG system within PyCharm, a Python IDE from JetBrains, and demonstrate that our dataset remains indicative of real user behavior.

To summarize, this paper makes the following contributions: 
\begin{itemize}
    \item We collect and publish a novel dataset that allows us to measure both online and offline metrics for each commit~(\Cref{sec:data-collection}). We verify that it remains indicative of real user behavior by employing logs from PyCharm, a Python IDE from JetBrains~(\Cref{sec:data-validation}).
    \item To help the community, we share the application for the dataset collection~(\Cref{sec:data-app}) and the way to extend the dataset synthetically while maintaining quality~(\Cref{sec:synth-data,sec:data-validation}).
    \item We describe the method for metric selection and share our findings~(\Cref{sec:metrics-study}).
    \item We release the code and the dataset to support future research in the field of CMG: \url{https://jb.gg/cmg-evaluation}.
\end{itemize}

\section{Framework}
\label{sec:framework}

\begin{figure}[h]
\centering
\includegraphics[width=\columnwidth]{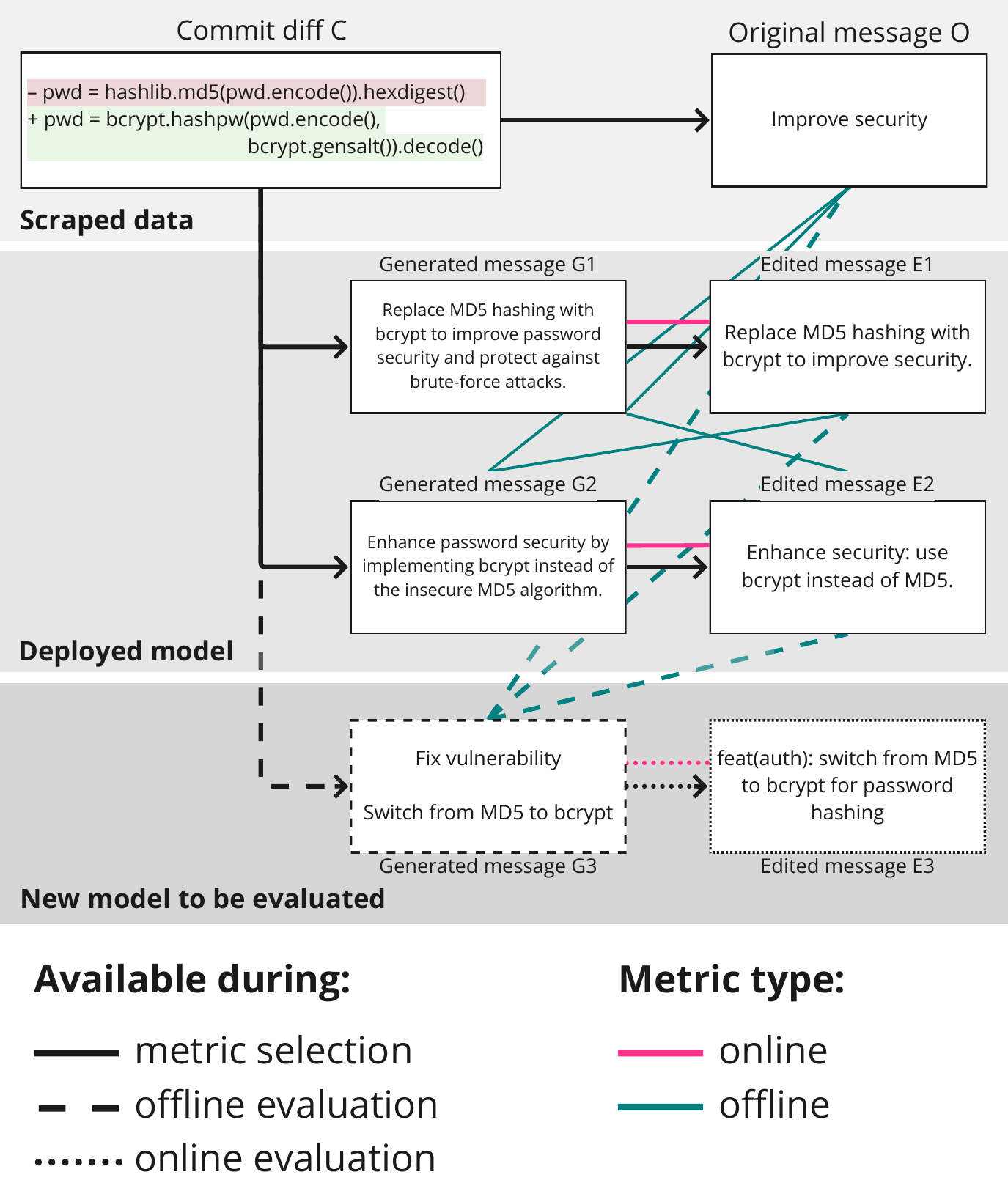}
\caption{We propose to collect multiple pairs of generated messages $G$ and their edited versions $E$ for each commit $C$ to find an \textcolor[HTML]{008080}{\textbf{offline}} metric (\textcolor[HTML]{008080}{\textbf{teal}} lines on the scheme) that estimates the \textcolor[HTML]{FF318C}{\textbf{online}} metric (\textcolor[HTML]{FF318C}{\textbf{magenta}} lines on the scheme) as best as possible. This allows us to select the model that will show the best quality when deployed and evaluated \textit{online} (dotted lines) while evaluating it in the \textit{offline} setting (dashed line).}

\label{fig:motivation}
\end{figure}

In this section, we explain the main concepts of our framework and how we envision its possible application. The process is illustrated in~\Cref{fig:motivation}. 

Imagine that there is an already deployed CMG system with available \textcolor[HTML]{FF318C}{\textbf{online}} metrics scores (the middle part of~\Cref{fig:motivation}). For any new version of a CMG model (the lower part of~\Cref{fig:motivation}), to compare it with the current version would require deploying it and collecting \textcolor[HTML]{FF318C}{\textbf{online}} metrics over a period of time, which is a lengthy process. Instead, we propose a framework to choose the metrics for an \textcolor[HTML]{008080}{\textbf{offline}} evaluation setup based on their correlation with the metrics from the \textcolor[HTML]{FF318C}{\textbf{online}} setup. In this way, we facilitate the fast experimentation offered by \textcolor[HTML]{008080}{\textbf{offline}} evaluation setup while ensuring that the results remain relevant to the real users' signals from the \textcolor[HTML]{FF318C}{\textbf{online}} evaluation setup. 

Most existing datasets for \textcolor[HTML]{008080}{\textbf{offline}} evaluation of CMG models are collected from open sources, \emph{e.g.}, GitHub. There are two key entities associated with each commit in the VCS: the set of code changes $C$ and an original commit message $O$ written by developers (the upper part of~\Cref{fig:motivation}). The quality of the CMG model is then obtained by calculating \textcolor[HTML]{008080}{$\mathbf{m(G,O)}$}, where \textcolor[HTML]{008080}{$\mathbf{m}$} is a text similarity metric, and $G$ is a model-generated commit message. However, during real usage of a CMG system, users often edit the generated message $G$ before submitting it to VCS, thus introducing the third entity for a commit---the edited message $E$. We illustrate this idea with motivating examples: for instance, a user may want to shorten the generated message (in~\Cref{fig:motivation}, $E_1$ and $E_2$ are shortened versions of $G_1$ and $G_2$, respectively) or to rewrite it to follow the conventions from their project (in~\Cref{fig:motivation}, $E_3$ is a version of $G_3$ adapted to the Conventional Commits~\cite{conventional-commits} specification). Among the \textcolor[HTML]{FF318C}{\textbf{online}} metrics, a common choice is to estimate the effort the user edit took (\emph{e.g.}, time, edit distance~\cite{levenshtein}). Guided by this intuition, we propose to consider the correlation of \textcolor[HTML]{008080}{$\mathbf{m(G,O)}$} with the estimation of the editing effort as the indication of how well metric \textcolor[HTML]{008080}{$\mathbf{m}$} approximates the way real users perceive quality. Specifically, at JetBrains, \textcolor[HTML]{FF318C}{$\mathbf{ED(G,E)}$}---edit distance~\cite{levenshtein} between the generated messages $G$ and their user-edited versions $E$---is one of the important indicators of online quality of our CMG system. Hence, in our experiments, we use \textcolor[HTML]{FF318C}{$\mathbf{ED(G,E)}$} as the online measure.

To calculate this correlation, we require a dataset that, unlike existing ones, includes not only code changes $C$ and the original commit message $O$ but also the model-generated messages $G$ and their edited versions $E$. After we select the best-performing \textcolor[HTML]{008080}{\textbf{offline}} metric \textcolor[HTML]{008080}{$\mathbf{m}$}, we can use \textcolor[HTML]{008080}{\textbf{offline}} evaluation to experiment with the new model (dashed lines in~\Cref{fig:motivation}) before deploying it into production and collecting actual \textcolor[HTML]{FF318C}{\textbf{online}} metrics (dotted lines in~\Cref{fig:motivation}). 

Moreover, we propose to collect several $(G, E)$ pairs for each commit. For each model-generated message $G$, there is a \textit{related} pair $E$ that was directly obtained from $G$. However, the edited versions of other generated messages can be considered \textit{conditionally independent} with $G$ given the commit $C$, just like the original commit message $O$. At the same time, all $E$ undergo additional verification, so these \textit{conditionally independent} references might even be of higher quality than an original commit message $O$ from open sources. In~\Cref{fig:motivation}, we show two different messages $G_1$ and $G_2$ generated by an already deployed CMG model, and their edited versions, $E_1$ and $E_2$, respectively. While $E_1$ is \textit{related} to $G_1$, and the distribution of $\mathbf{m(G_1,E_1)}$ is likely to differ from the usual \textcolor[HTML]{008080}{$\mathbf{m(G_1, O)}$}, $E_2$ is \textit{conditionally independent} given the commit $C$, and computing $\mathbf{m(G_1,E_2)}$ is just as valid as \textcolor[HTML]{008080}{$\mathbf{m(G_1, O)}$}. This allows us to estimate the quality of the same message $G_1$ in both ways, online (versus $E_1$) and offline (versus $E_2, O$).

\section{Dataset}
\label{sec:dataset}

\begin{figure*}[h!]
\centering
\subfloat[Expert labeling.]{
    \includegraphics[width=0.9\textwidth]{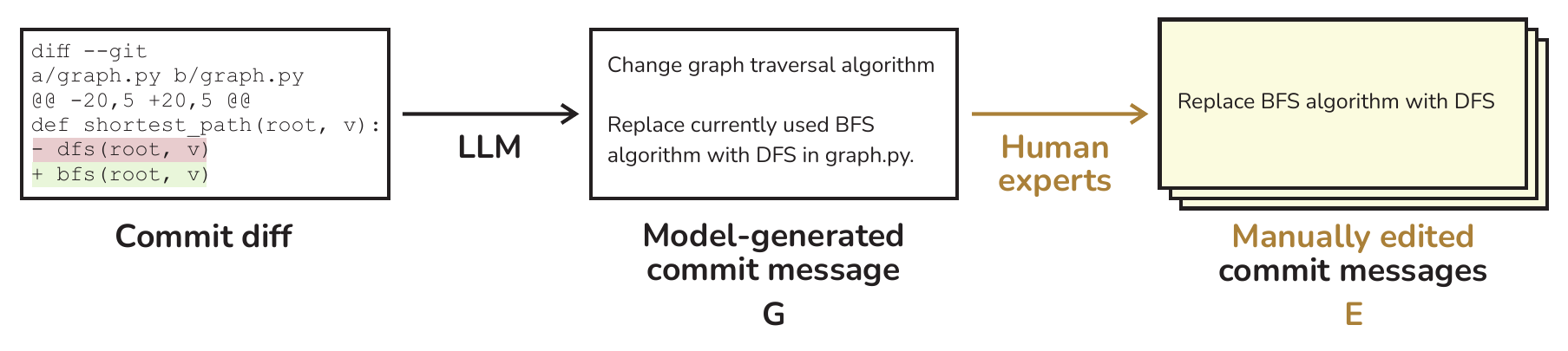}
    \label{fig:dataset-human-edit}
}

\subfloat[Synthetic backward generation.]{
    \includegraphics[width=0.9\linewidth]{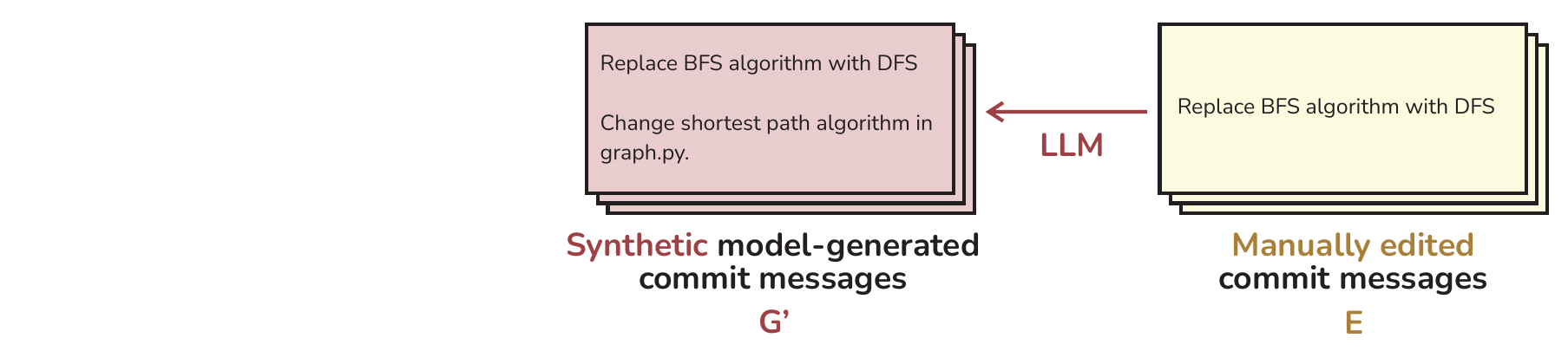}
    \label{fig:synthetic-backward}
}

\subfloat[Synthetic forward generation.]{
    \includegraphics[width=0.9\linewidth]{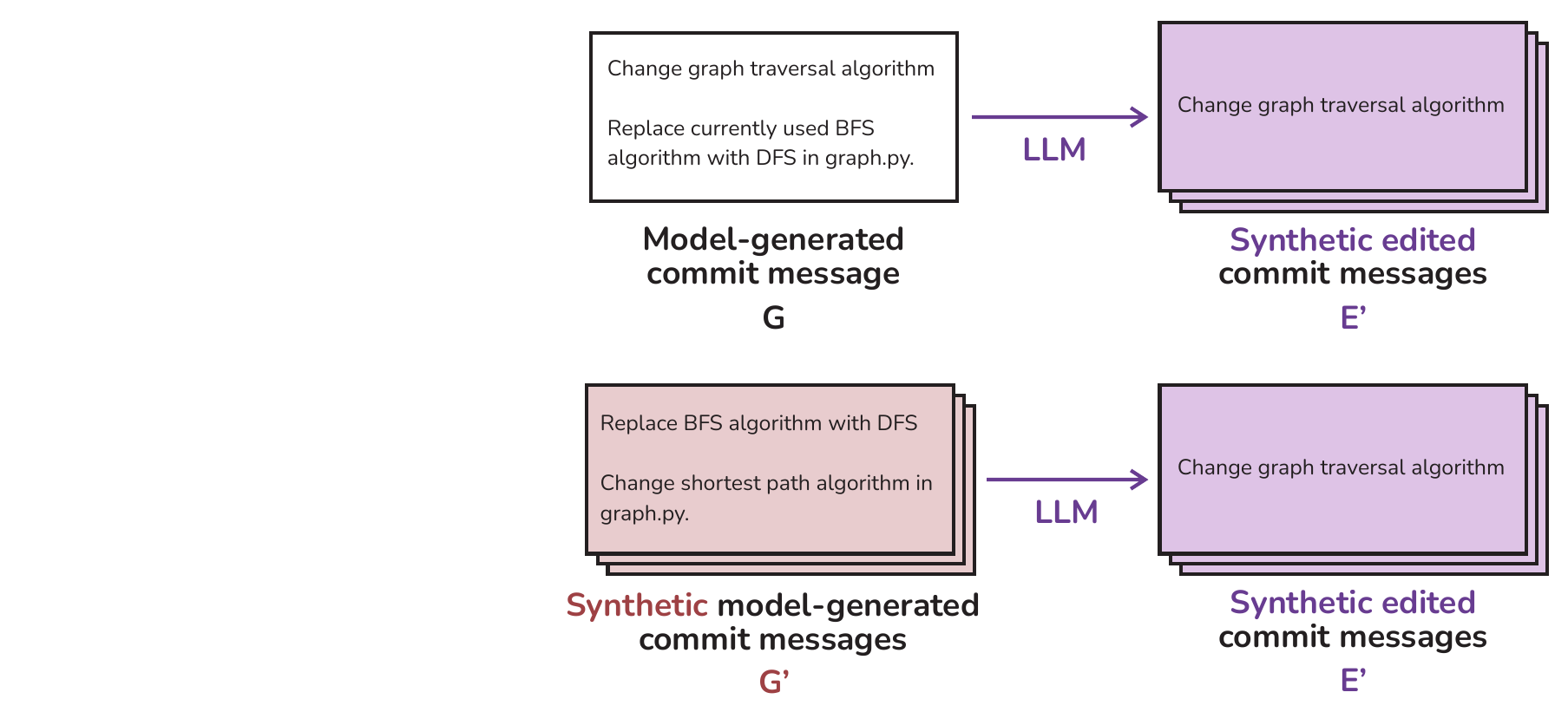}
    \label{fig:synthetic-forward}
}
\caption{Overview of our dataset collection process.}
\label{fig:dataset-collection}
\end{figure*}

To enable the framework described in~\Cref{sec:framework}, we require a dataset with the following information for each commit: $C$, the code changes to be committed; $O$, the original commit message from VCS; $G$, model-generated commit messages; and $E$, the edited versions of messages $G$. From practical considerations, we use \textcolor[HTML]{FF318C}{$\mathbf{ED(G,E)}$} as an estimation of the user effort the editing took for each pair $(G,E)$, so no additional data is required. Since there are no existing datasets providing generated and edited messages, in this work, we build a novel dataset with multiple $(G,E)$ pairs per commit.

The process of dataset collection is illustrated in~\Cref{fig:dataset-collection}. We implemented a web application (\Cref{sec:data-app}) for labeling and collecting  message edits \textcolor[HTML]{c19c0b}{\textbf{from human experts}} (\Cref{sec:data-collection}, illustrated by~\Cref{fig:dataset-human-edit}). Given the manual effort of rewriting commit messages and the high quality of synthetic data generation reported in numerous research works~\cite{synthetic-survey}, we then propose two different ways to synthetically augment the dataset, \textcolor[HTML]{913632}{\textbf{backward}} and \textcolor[HTML]{58136a}{\textbf{forward}} (\Cref{sec:synth-data}, illustrated by \Cref{fig:synthetic-backward} and \Cref{fig:synthetic-forward}, respectively). Finally, we show that our dataset is indicative of real user behavior by comparing the distribution of edits with user logs from PyCharm, a Python IDE developed by JetBrains (\Cref{sec:data-validation}).

\subsection{Dataset Collection Application}\label{sec:data-app}

\begin{figure*}[h!]
\centering
\includegraphics[width=0.95\textwidth]{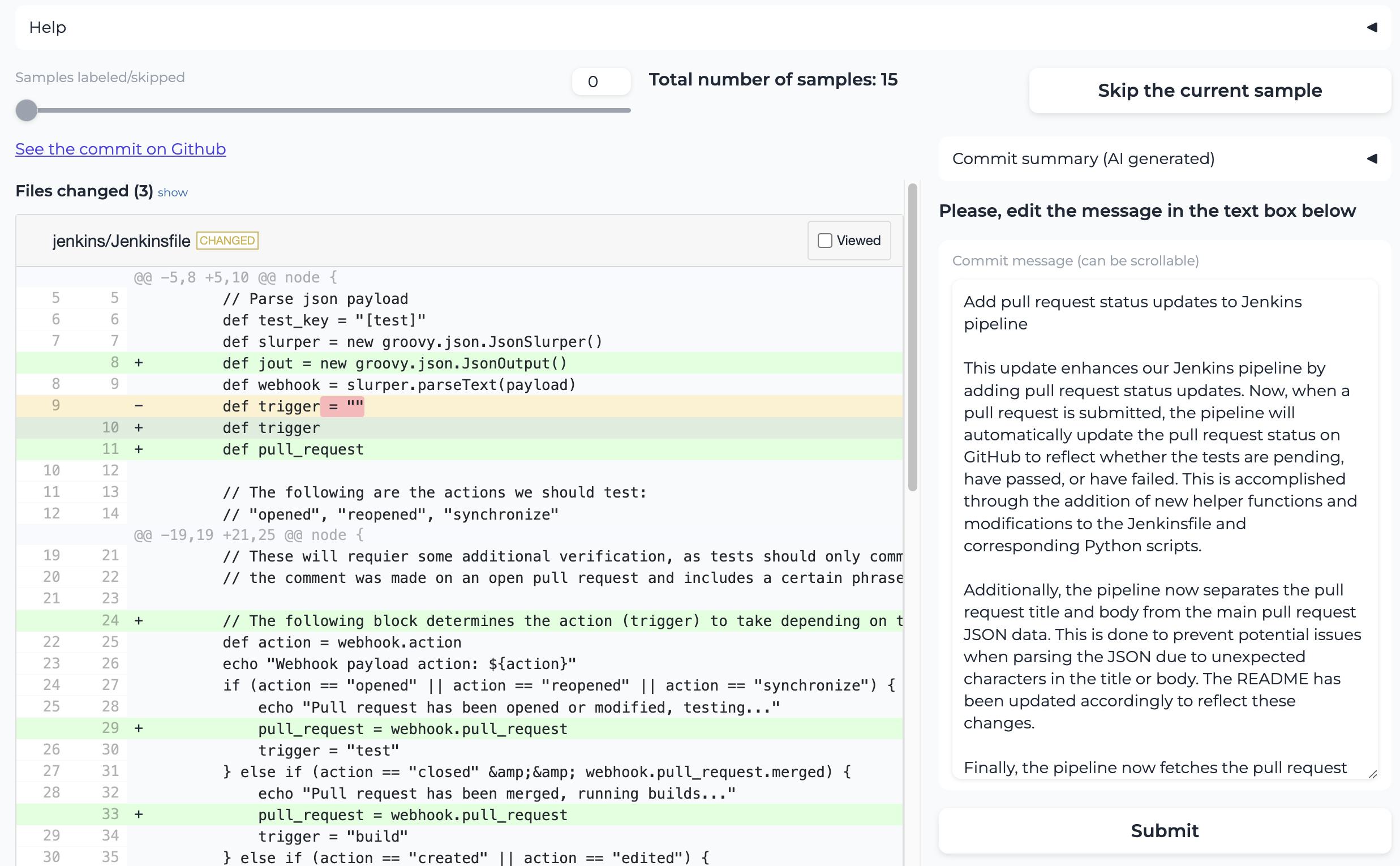}
\caption{A screenshot of our web application for the collection of commit message edits. On the left, the assessors are presented with a set of code changes from the current commit in a diff format. On the right, there is a window with a model-generated commit message to be edited, all changes in this window are tracked. Additionally, there is a \textit{Help} toggle with a labeling instruction on top and a \textit{Commit summary} toggle with extra information about the current project, aimed to provide more context for the code changes. The web application is publicly available~\cite{cmg-web-application}.}
\label{fig:collection-app-screenshot}
\end{figure*}

JetBrains values the privacy of its users and thoroughly follows privacy regulations, such as the General Data Protection Regulation (GDPR), and thus does not store user's data like the final edited messages $E$ that get submitted to the VCS.

So, to collect a dataset closely simulating the intended user behavior, we prepare and release to the open-source a web application~\cite{cmg-web-application} implemented using the Gradio framework~\cite{Abid_Gradio_Hassle-free_sharing_2019}. The graphical interface of the application is presented in~\Cref{fig:collection-app-screenshot}. The application provides an assessor with the diff---the set of code changes from the commit---and a window with the model-generated message $G$ to be edited. The application collects all the keystrokes in the editing window, including content, location and timestamp for each change, so the user typing can be fully reproduced from the collected data.

Since uncovering the rationale behind the code changes requires a deep understanding of the codebase, we add an LLM-generated summary of the whole project and an extended description of the intended changes behind the patch to aid the assessors. 
We strictly warn assessors not to copy-paste text from the extended description.

\subsection{Dataset Collection}\label{sec:data-collection}

The labeling was carried out by eight researchers from JetBrains Research and three PhD students with several years of experience in the industry. All participants were given the same set of commits, but the commits were randomly shuffled to avoid possible bias from the predefined order. 

As the source of the commits, we selected the Commit Message Generation benchmark from Long Code Arena~\cite{lca}. It features $163$ commits from Python repositories, all of which are manually confirmed to have complex and meaningful code changes akin to those developers encounter in their daily work. For model-generated commit messages $G$ that experts need to edit, we employed the results from GPT-4 (\texttt{gpt4-06-13}), a powerful proprietary LLM from OpenAI~\cite{gpt4}, released by the authors of the benchmark~\cite{lca-results}. The prompt and the code used to obtain the GPT-4-generated messages are available in the replication package of Long Code Arena~\cite{lca-repl}. 

We randomly selected a single set of $15$ commits that we present to each participant to achieve a higher number of labels per sample. 
We argue that having multiple labels per sample offers significant advantages. For each model-generated message $G$, all edited messages except for the related one can serve as high-quality references. Drawing inspiration from classical metrics used in reference-based evaluation, such as BLEU~\cite{Papineni02bleu:a} or ROUGE~\cite{lin-2004-rouge}, we believe that employing multiple references can lead to more stable and reliable evaluations. Moreover, we intentionally presented the labelers with an abstract instruction of ``edit the commit message so that it is good enough to publish to VCS'' to collect different types of edits, so having only one edited sample per commit would have inadvertently introduced a bias. 

In total, we collected $57$ \textcolor[HTML]{c19c0b}{\textbf{manually edited}} samples, with $3.8$ samples per commit on average. Our \textcolor[HTML]{c19c0b}{\textbf{expert-labeled}} dataset is available online~\cite{cmg-expert-dataset}.

\begin{table*}
\centering
\caption{Summary of the collected dataset.}
\resizebox{0.9\textwidth}{!}{
\begin{tabular}{ccccc}
\toprule
\textbf{Source} & \makecell{\textbf{Number of}\\\textbf{related pairs}} & \makecell{\textbf{Avg number of}\\\textbf{related pairs} \\\textbf{per commit}} & \makecell{\textbf{Number of}\\\textbf{conditionally}\\\textbf{independent}\\\textbf{pairs}} & \makecell{\textbf{Avg number of}\\\textbf{conditionally}\\\textbf{independent}\\\textbf{pairs per commit}}\\
\midrule
\textcolor[HTML]{c19c0b}{\textbf{Expert Labeling}} 
& 57 & 3.8 & --- & ---\\
\addlinespace[0.5ex]
\textcolor[HTML]{913632}{\textbf{Synthetic (backward)}} 
& 104 & 7.43 & 1048 & 74.86\\
\addlinespace[0.5ex]
\makecell{\textcolor[HTML]{58136a}{\textbf{Synthetic (forward)}}\\\textbf{from} \textcolor[HTML]{c19c0b}{\textbf{expert-labeled}}} 
& 177 & 11.8 & --- & ---\\
\addlinespace[0.5ex]
\makecell{\textcolor[HTML]{58136a}{\textbf{Synthetic (forward)}}\\\textbf{from} \textcolor[HTML]{913632}{\textbf{backward}}} & 318 & 22.71 & 3753 & 268.07\\
\midrule
\textbf{Full} & 656 & 43.73 & 5140 & 342.67\\
\bottomrule
\end{tabular}
}
\label{tab:dataset-summary}
\end{table*}

\subsection{Dataset Extension}\label{sec:synth-data}

Manually editing commit messages takes significant effort. Since we intend our framework to be useful for practitioners, it is important to maintain resource efficiency. With that in mind, we explore the possibility of synthetically extending our dataset~\cite{synthetic-survey}. To ensure this approach does not introduce bias, we compare the distribution of the synthetic subset with real interaction logs from PyCharm, a Python IDE developed by JetBrains, as detailed in Section~\ref{sec:data-validation}.

To imitate human actions, we employ GPT-4 Turbo~\cite{gpt4turbo} (\texttt{gpt-4-1106-preview}), a powerful LLM from OpenAI, together with the \textit{in-context learning (ICL)} technique. Specifically, we propose two different approaches to extend our dataset, illustrated in \Cref{fig:synthetic-backward,fig:synthetic-forward} and detailed below. We release the code used to obtain the synthetic data~\cite{cmg-code}.

For \textcolor[HTML]{913632}{\textbf{backward generation}} (\Cref{fig:synthetic-backward}), LLM is expected to ``invert'' the edit performed by a human expert. In other words, the LLM is given the manually edited version of a commit message $E$, and its task is to ``restore'' the original model-generated message $G$. To aid generation, we employ in-context learning and supply the model with several examples of pairs $E,G$ for different commits. We denote the result of such backward generation $G'$---a synthetic model-generated commit message. 
To eliminate the samples that are too far from the edit distributions from the expert-labeled dataset, we discard samples that have more than $T_b\%$ of content added during this step, where $T_b$ is a hyperparameter to be set.

For \textcolor[HTML]{58136a}{\textbf{forward generation}} (\Cref{fig:synthetic-forward}), LLM imitates precisely what human experts did during labeling, \emph{i.e.}, rewrites a model-generated commit message (both usual $G$ and synthetic $G'$)  to $E$. Here, as ICL examples of such conversion, we again used pairs of $G,E$ for different commits. The result is denoted $E'$---a synthetic edited commit message.
To alleviate the samples that are too far from the edit distributions from the expert-labeled dataset, we discard samples that have more than $T_f\%$ of content removed during this step, where $T_f$ is a hyperparameter to be set.

We query the LLM three times per message, allowing it up to three attempts to pass the set thresholds. We set the number of ICL samples to $15$ and the thresholds $T_b$ and $T_f$ to $50\%$ and $75\%$, respectively, based on our preliminary experiments. We leave a thorough search for the optimal parameters to be explored in the future work.

\begin{figure*}
\centering
\includegraphics[width=1\textwidth]{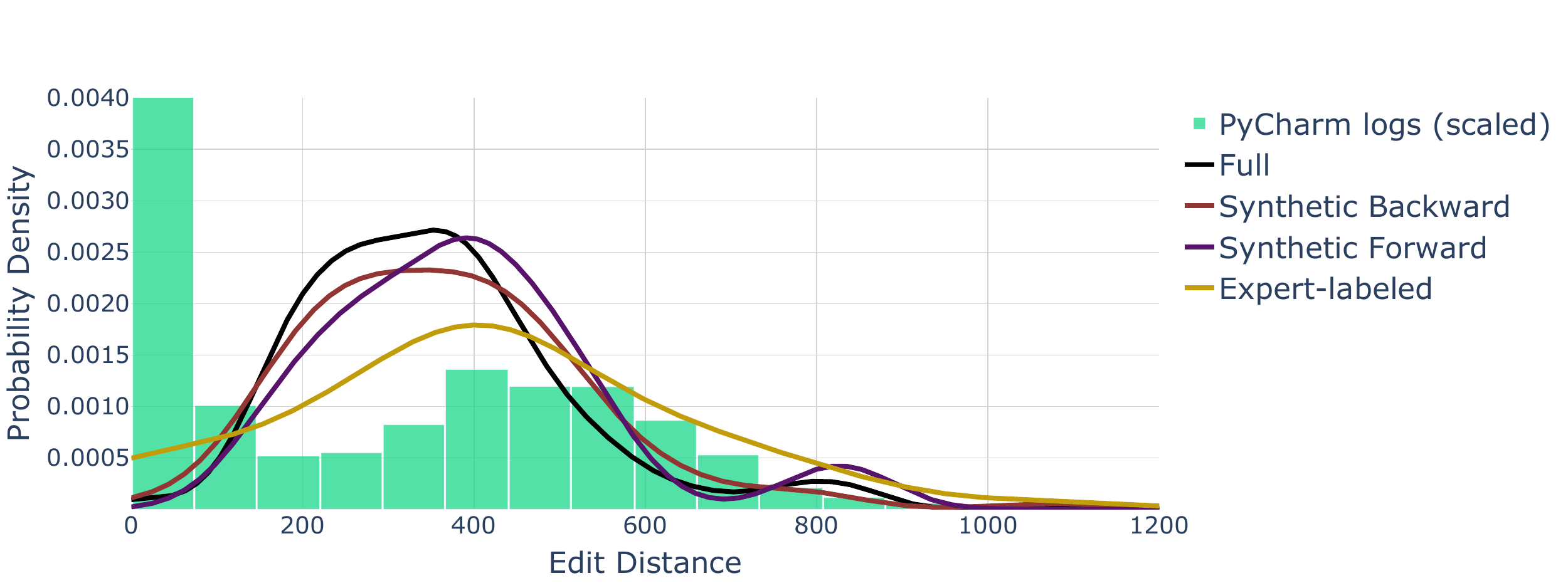}
\caption{Distribution of the $\mathbf{ED(G,E)}$ values for the different subsets of our dataset and for the PyCharm users logs over one month in April-May 2024, where $\mathbf{ED(G,E)}$ is the edit distance between model-generated messages $\mathbf{G}$ and their edited versions $\mathbf{E}$. Note that we scale the PyCharm logs $\mathbf{ED}$ values to adjust for the differences in messages' lengths and discard the samples with edit distance equal to 0.}
\vspace{-0.2cm}
\label{fig:fus-editdist-dist}
\end{figure*}

 The detailed statistics for the resulting dataset are provided in~\Cref{tab:dataset-summary}. Our \textbf{synthetic extension} of the dataset is also available online~\cite{cmg-synthetic-dataset}.

\subsection{Data Validation}\label{sec:data-validation}

As part of the Early Access Program for IntelliJ-based IDEs, JetBrains collects anonymous usage statistics. This includes \textcolor[HTML]{FF318C}{$\mathbf{ED(G,E)}$} --- the edit distance between the commit message $G$ generated by the AI Assistant plugin and the message $E$ that actually gets stored in the VCS. To ensure that our expert-labeled dataset and its synthetic extension remain representative of real usage, we compare the distributions of \textcolor[HTML]{FF318C}{$\mathbf{ED(G,E)}$} in our dataset to the same metrics computed for actual users of PyCharm, a Python IDE developed by JetBrains.

Figure~\ref{fig:fus-editdist-dist} shows the distribution of edit distance values for the PyCharm users over one month in April-May 2024, as well as for the different subsets of our dataset. 
We note that the majority of the samples in PyCharm user logs have very low edit distance values ($78\%$ of all samples have an edit distance value equal to $0$), while in our dataset, there are no such samples at all. One possible reason is that in such cases, the quality of the generated commit messages is sufficient for the users to keep them as is. On the other hand, commit messages might not be as important compared to the effort required to verify and edit them. Though there are a lot of samples with edit distance equal to $0$, they come from only $33$\% of users. As we do not want to consider the cases with low user engagement, we remove the samples with a zero edit distance, leaving us with $22$\% of total samples coming from $67$\% of total users.

Additionally, we observe that the average length of generated commit messages $G$ in our dataset and in PyCharm user logs varies, causing a shift in edit distance distribution, as edit distance does not take the length of the text into account. To adjust for this, we multiply the PyCharm logs' edit distance by the coefficient $R \approx 1.77$, the ratio of the average length of model-generated commit messages in our dataset to those from the user logs.

As observed in~\Cref{fig:fus-editdist-dist}, real PyCharm users tend to perform less editing than experts in our dataset. However, we note that the peak of the edit distance distribution, $\approx 400$, for the user logs data is relatively close to the peak for the golden expert-labeled dataset. Therefore, the data we collected is, to a certain extent, representative of real-life user behavior.

As for the synthetic extension of our dataset, we observe that, out of all subsets, the one obtained via \textit{backward generation} is the closest to the expert-labeled subset. Nevertheless, the full dataset also remains representative. We use the full dataset for subsequent experiments, as it contains the largest number of pairs.

\section{Metrics Study}\label{sec:metrics-study}

\begin{figure}
\centering
\includegraphics[width=\linewidth]{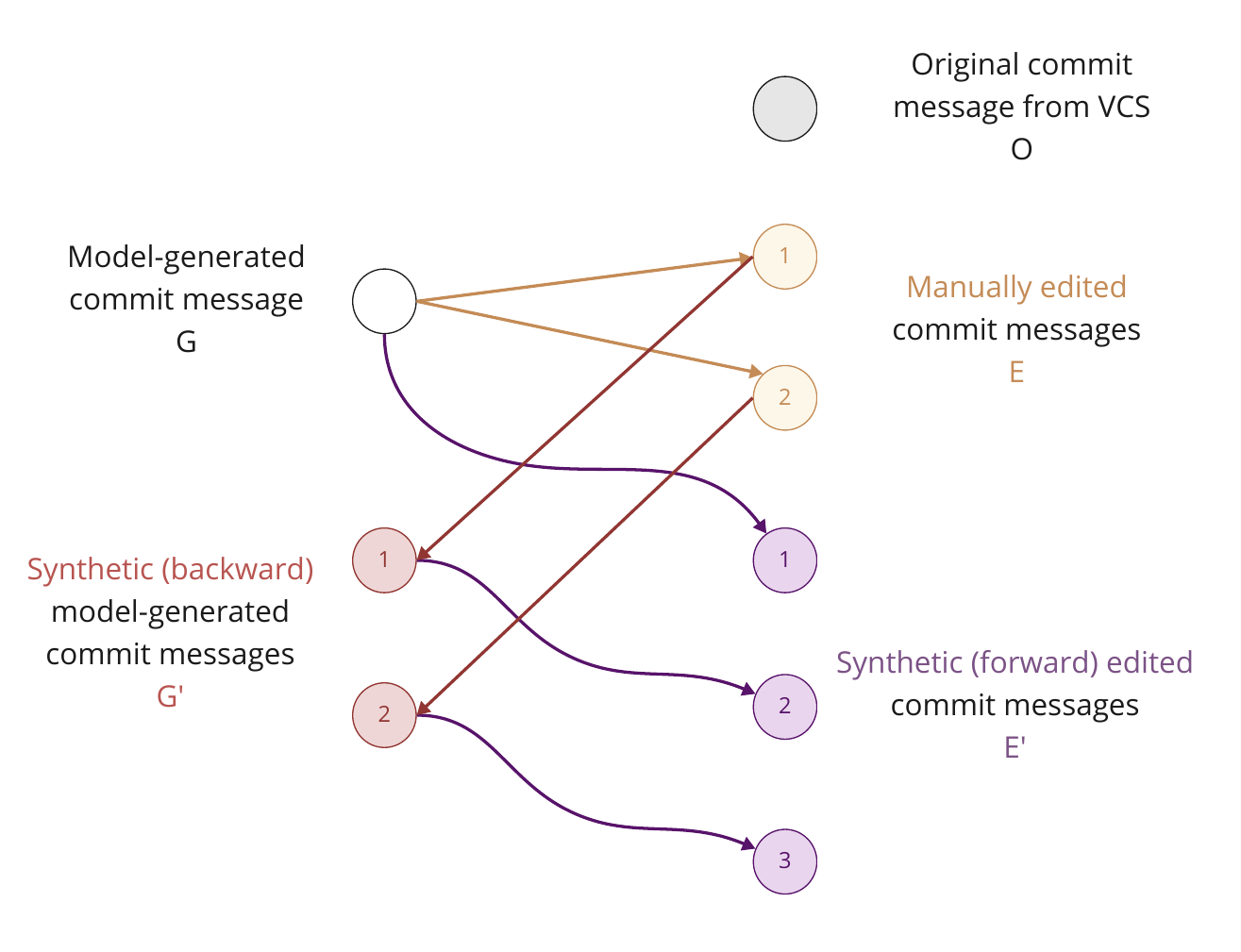}
\caption{Overview of the relations between different $(G,E)$ pairs in our dataset. The pairs connected with the lines are the related pairs. Any pair $(G,E)$ for a given commit message $C$ that is not related is \emph{conditionally independent}. We employ \textit{related} pairs to calculate online metrics and \textit{conditionally independent} pairs to calculate offline metrics.}
\vspace{-0.2cm}
\label{fig:metrics-overview}
\end{figure}

\subsection{Motivation}

In the dataset that we collected and presented in Section~\ref{sec:dataset}, for each set of code changes to be committed $C$, we have multiple pairs of model-generated commit messages $G$ and their edited counterparts $E$. 
For each model-generated message $G_i$, some of the edited messages are \textit{related} (\emph{e.g.}, if $E_i$ is a manually edited version of $G_i$, or if $G_i$ and $E_i$ were obtained from one another synthetically), and the rest of edited messages are \textit{conditionally independent} given $C$ and can be viewed as additional high-quality references on top of $O$, the original commit message for the current commit from VCS. \Cref{fig:metrics-overview} presents all possible kinds of messages (\textcolor[HTML]{808080}{\textbf{original}}, \textcolor[HTML]{c19c0b}{\textbf{expert-edited}}, \textcolor[HTML]{913632}{\textbf{synthetic backward}}, \textcolor[HTML]{58136a}{\textbf{synthetic forward}}) and the direct \textit{relations} between them, following the data collection process described in~\Cref{sec:dataset}.

This way, for each commit, we obtain both $ED(G_i,E_j), E_j \in \text{related}(G_i)$, and $m(G_i,E_j), E_j \in \text{conditionally independent}(G_i)$, where $m$ is a text similarity metric and $ED$ is the edit distance metric that we employ as an estimation of the user effort the edit took, as it is one of the important online metrics computed for the actual CMG system at JetBrains. We do not use the time required to perform an edit because it is available only for the expert-labeled part of our dataset and not for the synthetic.

The $ED(G_i,E_j), E_j \in \text{related}(G_i)$ values better describe the quality of the CMG system from the point of view of its end users. However, obtaining them involves manually editing model-generated messages, so the process is infeasible to scale to large datasets or repeat numerous times, in contrast with reference-based evaluation with automatic metrics. Instead, we propose to use these values to select the most suitable metrics $m$ for a reference-based evaluation. Put another way, our core assumption is that the correlation of $m$ with $ED(G_i,E_j), E_j \in \text{related}(G_i)$ is indicative of how good $m$ is. Additionally, following the design of many classical metrics, \emph{e.g.}, BLEU~\cite{Papineni02bleu:a} or ROUGE~\cite{lin-2004-rouge}, we use the multiple references in our dataset and aggregate $m$ values. We argue that the conditionally independent messages $E$ might serve as references of high quality for each message $G$ as compared to the original commit message $O$, considering issues with the quality of commit messages coming from the open sources~\cite{good-commit-message}.

\subsection{Method}

For each commit in our dataset, we compute $ED$ for all related pairs $(G_i, E_j)$ where $E_j \in \text{related}(G_i)$, \emph{i.e.}, it is an edited version of $G_i$. When $|\text{related}(G_i)| > 1$, \emph{i.e.}, there are more than one edited version of $G_i$ available, we take the average value over $\text{related}(G_i)$.

On the other hand, we compute $m$ for all conditionally independent pairs $(G_i, E_j)$, where $E_j \in \text{conditionally independent}(G_i)$, \emph{i.e.}, it is an edited version of some other message $G_j$ and is not directly related to $G_i$. As an aggregation function $f_{agg}$, similarly to BLEU and ROUGE, we use the maximum if the higher values of $m$ indicate better quality or the minimum if the lower values of $m$ indicate better quality, \emph{e.g.}, for edit distance.
Finally, we compute the correlation between these two calculated values: 
\begin{equation}
\begin{split}
Q(m) := \text{corr } \Big( \text{mean}\big(\{ED(G_i, E_j)\}_{E_j \in \text{related}(G_i)}\big), \\ 
f_{agg}\big(\{m(G_i, E_j)\}_{E_j \in {\text{conditionally independent}(G_i)}}\big) \Big)
\end{split}\label{eq:metrics-corr}
\end{equation}

We use the Spearman rank correlation coefficient from the SciPy package~\cite{scipy} in our analysis. In~\Cref{eq:metrics-corr} describing our proposed framework, as the text similarity metric $m$, we consider the following candidate metrics: BLEU~\cite{Papineni02bleu:a} (using the variation that showed the highest correlation with human judgement for CMG~\cite{tao2021evaluation}), ROUGE~\cite{lin-2004-rouge}, METEOR~\cite{banarjee2005}, BERTScore~\cite{bert-score}, chrF~\cite{popovic-2015-chrf}, edit distance~\cite{levenshtein} and edit similarity (which is the edit distance normalized by the maximum length among prediction and reference). 

BLEU, ROUGE, and METEOR are classical metrics from the Natural Language Processing (NLP) field, which have been widely applied to numerous tasks, including CMG, as previous studies indicate~\cite{tao2021evaluation, zhang2024automatic}. ChrF and BERTScore are also renowned for various NLP tasks, though they have not been applied to CMG previously. Similarly, edit distance is rarely used in CMG evaluation; however, edit similarity was applied to CMG before~\cite{eliseeva2023commit}. All of these metrics except for BERTScore are based on the word-level or character-level overlap between predictions and references, while BERTScore employs a language model to capture the semantic aspects.

Note that for most of the selected metrics, larger values suggest a closer match between the prediction and the reference. However, for the online metric, edit distance, larger values indicate greater differences between the two strings. Therefore, negative correlations are expected for all metrics except edit distance.

Lastly, the primary online metric we consider is the edit distance between generated messages and their edited counterparts. However, we acknowledge that other metrics could be a valid choice for online evaluation as well. Our dataset and framework allows to easily replace $ED$ in $ED(G_i, E_j)$ where $E_j \in \text{related}(G_i)$ with any other metric $m^*$. In addition to our primary experiments with $Q(m)$, we also consider the correlation coefficients:
\begin{equation*}
\begin{split}
Q^*(m, \textcolor{red}{m^*}) := \text{corr } \Big( \text{mean}\big(\{\textcolor{red}{m^*}(G_i, E_j)\}_{E_j \in \text{related}(G_i)}\big), \\ 
f_{agg}\big(\{m(G_i, E_j)\}_{E_j \in {\text{conditionally independent}(G_i)}}\big) \Big)
\end{split}\label{eq:metrics-corr-extra}
\end{equation*}

We compute and release $Q^*(m, m^*)$ for the set of 9 text similarity metrics listed above used both as an offline metric $m$ and as an online metric $m^*$~\cite{cmg-code}.

\subsection{Results \& Discussion}

\begin{table}
\centering
\caption{Spearman correlation coefficients $Q(m)$ as defined in \Cref{eq:metrics-corr}. The metrics are sorted by the absolute values of correlation coefficients. Note that our online metric is edit distance, where larger values indicate greater differences. For all considered metrics $m$ (except edit distance), it is not the case, hence, \textit{negative} correlation is expected.}

\begin{tabular}{c|c|c|c}
\toprule

& \makecell{Metric\\$m$} & \makecell{Correlation\\$Q(m)$} & p-value\\
\midrule
\textbf{High Correlation} & Edit Distance & 0.74 & $< 0.05$\\
\midrule
\textbf{Moderate Correlation} & Edit Similarity & -0.36 & $< 0.05$\\
\midrule
\multirow{7}{*}{\textbf{Low Correlation}} 
& BERTScore & -0.26 & $< 0.05$\\
& ROUGE-L & -0.26 & $< 0.05$\\
& ROUGE-2 & -0.20 & $< 0.05$\\
& ROUGE-1 & -0.19 & $< 0.05$\\
& BLEU & -0.17 & 0.07\\
& ChrF & 0.05 & 0.62\\
& METEOR & 0.04 & 0.66\\
\bottomrule
\end{tabular}

\label{tab:metric-corr}
\end{table}

We present the experimental results in~\Cref{tab:metric-corr}. We divide the text similarity metrics into three groups based on the obtained correlation coefficients $Q(m)$.

\paragraph{High correlation $(0.7 \leq abs(Q(m)) < 1.0)$} The only metric $m$ showing high correlation with the values $ED(G_i,E_i)$ that represent the online quality of the CMG system is edit distance. We hypothesize that the reason for edit distance in an offline evaluation setting having the highest correlation with edit distance in an online evaluation setting lies in the model-generated messages $G$ in our dataset being notably more verbose than both their edited versions $E$ and original messages $O$. In an expert-labeled subset of our dataset, the average number of characters for $G$ is $636$, while for $E$ and $O$, it is $282$ and $204$, respectively.

\lessonlearned{\textbf{Observation 1.} Edit distance shows a high correlation with the ``online'' quality of the CMG system, which makes it a good choice for reference-based CMG evaluation when a sufficient difference in lengths between model-generated and reference messages is expected.}

\paragraph{Moderate correlation $(0.3 \leq abs(Q(m)) < 0.7)$} In a moderately correlated group, we observe only edit similarity. As edit similarity is a normalized version of edit distance, similar reasoning applies: likely, this metric correlates with $ED(G_i,E_j), E_j \in \text{related}(G_i)$ due to model-generated messages $G$ being much longer than both $E$ and $O$ in our dataset.

\lessonlearned{\textbf{Observation 2.} Edit similarity is moderately correlated with the online quality of the CMG system. Like edit distance, edit similarity is suitable for cases when model-generated commit messages are significantly longer than considered references, though edit distance might be a better choice since it shows a higher correlation.}

\paragraph{Low correlation $(0.0 \leq abs(Q(m)) < 0.3)$} 
Surprisingly, we observe rather low correlations for widely used metrics for reference-based evaluation in CMG research works, such as BLEU, METEOR, and ROUGE. ChrF and BERTScore, which were not often applied to CMG, are also in this group.
Another finding is that, while larger values indicate closer similarity for all the metrics in this group (unlike our primary online metric, edit distance, for which the smaller values are better), we still observe a positive correlation with $ED(G_i,E_j), E_j \in \text{related}(G_i)$ for ChrF and METEOR in Table~\ref{tab:metric-corr}. However, for BLEU, ChrF, and METEOR, our findings are not statistically significant at the confidence level of 95\%. Overall, the applicability of metrics from this group remains unclear, as the results are mixed.

\lessonlearned{\textbf{Observation 3.} BERTScore, ROUGE, BLEU, ChrF, and METEOR show low correlation with the online quality of the CMG system and thus might not be the best choice for reference-based CMG evaluation.}

\section{Limitations}

A key limitation of our study is that we resort to building a dataset of generated commit messages $G$ and their edited versions $E$ instead of employing data produced by real users of a CMG system deployed in production. However, this choice is motivated by privacy concerns, and we show that the distribution of the dataset we obtained is close to the one from PyCharm users.

Another limitation is that we conducted experiments for a single CMG model, a proprietary LLM GPT-4, in a zero-shot setup. Our key priority in model selection was to follow the up-to-date setup from the CMG system in the JetBrains AI Assistant product.
The findings we obtained may not transfer to CMG systems producing significantly different commit messages. For example, models generating less verbose messages might yield different results. Still, GPT-4 is a powerful LLM with strong performance across numerous tasks, including commit message generation~\cite{lca}, making it potentially representative beyond the specific JetBrains feature.

Next, we propose the framework for selecting the most suitable text similarity metric for offline experiments based on available logs before deploying a new version of the model. To further verify the framework, the next step would be to deploy the new version of the model and confirm that the findings from offline experiments remain relevant to the new logs from real users. Deployment to production and conducting online experiments are time-consuming processes and thus fall outside the scope of this work.

Finally, we note that the expert-labeled dataset we collected is relatively small, with only 57 manually edited samples. To be able to make reliable conclusions, we synthetically extend the dataset via an LLM. While synthetically generated data may not be perfect, it is a common step employed by numerous researchers and practitioners~\cite{synthetic-survey}. Additionally, we verify that the distribution of synthetically obtained samples remains close to both the expert-labeled dataset and the PyCharm user logs. Moreover, we describe our pipeline in detail and make all the artifacts publicly available to facilitate further research.

\section{Related Work}

Evaluation is a crucial aspect of both research works and production systems. Common approaches to evaluation include \textit{human evaluation} (\emph{i.e.}, evaluating the quality based on human experts' scores on a selected dataset)~\cite{chang2024survey}, \textit{offline evaluation} (\emph{i.e.}, evaluating the quality based on the automatic metrics on a selected dataset)~\cite{chang2024survey}, and, for production systems, \textit{online evaluation} (\emph{i.e.}, evaluating the quality based on logs of users' interactions with the system).

For the Commit Message Generation, specifically, previous studies focused on estimating the correlation of automatic metrics for reference-based offline evaluation with human evaluation. For instance, Tao et al.~\cite{tao2021evaluation} considered correlation with human scores on a 5-point Likert scale for three variations of the BLEU~\cite{Papineni02bleu:a} metric appearing in research works. Later, Hu et al.~\cite{hu2022correlating} estimated the correlation of several automatic metrics with human scores on a 5-point Likert scale for several criteria (naturalness, expressiveness, and more).

There are also numerous works that show how diverse the perception of commit message quality might be. Tian et al.~\cite{good-commit-message} conducted a multi-method study, introducing the taxonomy of commit messages based on the ``what'' (summary of the changes) and ``why'' (the reasons for the changes) information. This line of work was continued by Li and Ahmed~\cite{li2023commit} with a fine-grained analysis of the data considered in the previous study and a better automatic commit message quality classifier. On the other hand, Faragó et al.~\cite{farago2023full} take not only semantic aspects but the adherence to well-established commit message conventions into account.

In practice, there are even more factors that influence the user's satisfaction with a CMG system. We argue that human evaluation with a question rubric derived from theoretical principles is not guaranteed to reflect the real users' decisions in a production setting. Compared to related work, ours is the first to propose the framework to employ the user signals to choose a metric for reference-based evaluation, thus enabling fast experimentation by using offline evaluation but ensuring that the results are still in alignment with the online evaluation.

\section{Conclusions}

In this paper, we propose a novel framework for the evaluation of CMG systems that allows to select the best offline metric with respect to the target online metric. This, in turn, enables fast \textit{offline} experimentation while ensuring that the results are indicative of expected \textit{online} performance. In this framework, we define the quality of the generated commit message as the inverse of the complexity of the task of editing it. This approach is inspired by the observation that, in practice, users often modify commit messages generated by a CMG system before committing them to a VCS. From a practical standpoint, it can be posited that an ideal CMG system should be reliable enough that users trust the generated message implicitly, to the extent that they accept it without review or modification. We argue that, compared with the traditional \textit{human evaluation} that presents the assessors with a fixed question rubric, our setup comes closer to how users of actual CMG systems judge their performance. 

Due to the limitations of the real world, we make a series of simplifications to enable the proposed framework. 
Since collecting the usage logs directly from the production system will breach privacy, we resort to collecting a dataset of commit message edits from expert assessors. Additionally, to lower the amount of effort required to build such a dataset and keep our framework practical, we propose two ways to extend it synthetically, backward generation and forward generation. Finally, we compute the correlation with the online quality for nine renowned text similarity metrics employed for reference-based offline evaluation. Our findings reveal that edit distance shows the highest correlation, presenting an interesting contrary case to the common belief in the superiority of the semantic-similarity metrics over literal similarity metrics.

We make all the tools for data collection and metrics analysis, as well as expert-labeled and synthetic datasets, publicly available: \url{https://jb.gg/cmg-evaluation}. The metrics study in our work focused solely on a single CMG model, GPT-4, and the findings may not generalize to models with significantly different output formats. However, we believe that the proposed framework, along with the published artifacts, provides a solid foundation for further experimentation.

\section*{Acknowledgments}

We are grateful to the Evaluation team in the AI Assistant project and Michele Conti for the initial work on CMG evaluation.
We thank Nikolay Palichkov for the implementation of metrics collection in the IDE plugin. We appreciate  Svetlana Zemlyanskaya for proofreading the draft and reviewing the Dataset section.

\bibliographystyle{IEEEtran}
\balance
\bibliography{IEEEabrv,main_references}

\end{document}